\documentclass[11pt,a4paper]{article}
\usepackage[hyperref]{acl2018}
\usepackage{times}
\usepackage{latexsym}
\usepackage{graphicx}
\usepackage{url}
\usepackage[font=small]{caption,subcaption}
\aclfinalcopy

\title{De-anonymization of authors through arXiv submissions during double-blind review}

\author{Homanga Bharadhwaj , Dylan Turpin , Animesh Garg , and Ashton Anderson \\
   Department of Computer Science,  University of Toronto \\
   Vector Institute, Toronto }

\date{}

\begin{document}
\maketitle
\begin{abstract}
In this paper, we investigate the effects of releasing arXiv preprints of papers that are undergoing a double-blind review process. In particular, we ask the following research question: What is the relation between de-anonymization of authors through arXiv preprints and acceptance of a research paper at a (nominally) double-blind venue? Under two conditions: papers that are released on arXiv before the review phase and papers that are not, we examine the correlation between the reputation of their authors with the review scores and acceptance decisions. By analyzing a dataset of ICLR 2020 and ICLR 2019 submissions (n=5050), we find statistically significant evidence of positive correlation between percentage acceptance and papers with high reputation released on arXiv. In order to understand this observed association better, we perform additional analyses based on self-specified confidence scores of reviewers and observe that less confident reviewers are  more likely to assign high review scores to papers with well known authors and low review scores to papers with less known authors, where reputation is quantified in terms of number of Google Scholar citations. We emphasize upfront that our results are purely correlational and we neither can nor intend to make any causal claims. A blog post accompanying the paper and our scraping code will be linked in the project website \url{https://sites.google.com/view/deanon-arxiv/home} \footnote{Correspondence to \texttt{homanga@cs.toronto.edu}}.
\end{abstract}

\section{Introduction}
The use of single-blind reviews (which obscure reviewer identities) and double-blind reviews (which obscure both reviewer and author identities) varies across fields. Where double-blind review processes are preferred, they are often justified in terms of fairness (to lesser-known authors and institutions, to gender equity etc.) and reduced bias~\cite{snodgrass2006single}. In so far as reviewers are prevented from inferring author characteristics by obscuring author identity, reviewers cannot discriminate based on those characteristics. Blinding may also be an attempt to promote more objective reviewing, ensuring papers are judged only on their scientific merit.  

Unfortunately, double-blinding measures are always imperfect. Withholding author names obscures their identity, but can not guarantee that reviewers will not find out who wrote the paper some other way. Some sources of de-blinding include publication of a pre-print prior to review, putting up the paper on the authors' webpages, publicizing the paper through social media platforms etc.

Previous work has studied the efficacy of blinding measures in review processes by having authors guess the identity of their reviewers and vice versa. Such studies, in a variety of disciplines, report success rates for blinding of 53\% to 73\%~\cite{snodgrass2006single} (i.e. in the worst case, 47\% guessed correctly). Even where author names were removed from titles, identifiable details were sometimes left in the paper body or acknowledgements section. In small fields, the choice of project alone could be enough to identify authors.

In this paper, we study one possible source of de-blinding in papers submitted to ICLR (International Conference on Learning Representations). arXiv lets anyone immediately publish a citable technical report to the web, without any peer review. When ICLR papers under review are published on arXiv during the review process, it is possible reviewers will see the preprint and discover the authors and their affiliations.


\section{Related Works}

A recent paper by Tomkins et al.~\cite{singledouble} addressed a similar research question with an experimental study of the differences between single-blind (reviewer names withheld) and double-blind (reviewer and author names withheld) review processes. The authors designed a randomized controlled trial within the review process of the 10th WSDM conference. They divided reviewers into two categories: one that had access to authors’ names and affiliations (single-blind) and another that did not have access to the author list (double-blind). The same group of papers were reviewed by both groups of reviewers. Analysis of the bidding process and review scores revealed that reviewers in the single-blind pool were significantly more likely to recommend acceptance of papers from \textit{famous} authors, \textit{top} companies and \textit{top} universities.

Other papers relevant to our work have investigated different settings of reviewer bias. For example, Link et al.~\cite{usnonus} investigated the existence of reviewer bias when reviewers were asked to review manuscripts from authors outside of their home countries. In particular, they ran a controlled experiment during the review process for the \textit{Gastroenterology} journal and found that reviewers in the US assigned significantly higher review scores to papers with authors from US institutions compared to papers with authors outside the US.

\section{Concrete Operationalization}

We operationalize our research question as follows: Does the acceptance rate of ICLR papers correlate more strongly with author h-index (and total citations) when the identity of authors has potentially been revealed to reviewers by an arXiv preprint either during or before the review process?

\textbf{Choice of data}
The International Conference on Learning Representations (ICLR) is an emerging conference focused on deep learning. ICLR uses the \href{https://openreview.net/}{OpenReview} platform for open peer review, so all  reviews are publicly available for analysis. We choose ICLR data (from 2019 and 2020), because it contains information about acceptances, rejections and reviews of all submitted papers - including author data and affiliations. We scraped data from a total of 5057 submissions, after ignoring papers that were desk rejected or withdrawn prior to decision. 

\textbf{Data collection setup}
To collect ICLR data, we modified an existing tool\footnote{https://github.com/shaohua0116/ICLR2020-OpenReviewData} to scrape paper metadata from the OpenReview platform. arXiv provides an API for bulk data access\footnote{https://arxiv.org/help/bulk\_data}. Based on the papers and respective author lists scraped from OpenReview, we searched for papers with the same author list on arXiv, and for papers whose preprint existed on arXiv, we noted down the \textit{first upload} timestamp. We did not query using the paper title because it often happens that papers uploaded on arXiv have different titles compared to the paper submitted for review. 

Although Google Scholar does not provide an API for programmatic data access, however there are existing tools for scraping like scholarly\footnote{https://github.com/OrganicIrradiation/scholarly}. We found a brief news article in Nature~\cite{Else2018} written by a researcher who spent months scraping data from Google Scholar, for lack of an official public facing API. Luckily, since we only needed h-indices and total citations for several thousand authors, we wrote a simple script based on scholarly's source code that uses BeautifulSoup. In order to avoid getting our IP blocked by Google Scholars, we ran this scraping code on a server with timeouts between successive queries. 

Finally, we manually inspected the collected data to ensure that the first upload timestamp we scraped from arxiv does indeed correspond to the paper submitted on OpenReview. In addition, we manually checked the list of authors to ensure that we have scraped the h-index and total citations of the author we intended to from Google Scholars (since multiple people on Google Scholars might have the same name). We defer automated checks for these to future work. 

\textbf{Measuring author reputation}
To operationalize our research question, we have to choose a reasonable quantitative measure of \textit{author reputation}. For the purpose of this study, we define two metrics for the reputation of an author: their h-index and their total citations as calculated by Google Scholar. If an author does not have a Google Scholar page, they are excluded from our analysis. Overall, we had 5030 papers for our analysis.

\begin{figure}
    \centering
    \includegraphics[width=\columnwidth]{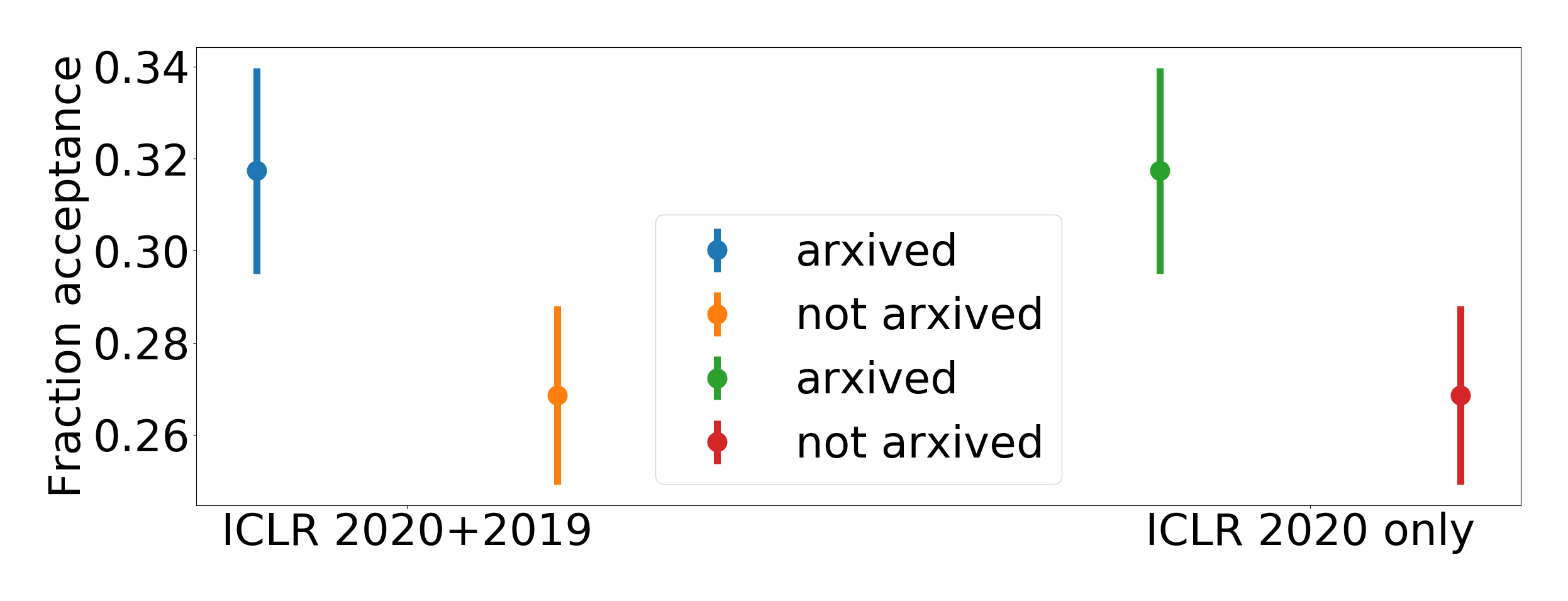}
    \caption{\%of papers released on arXiv for the \textit{arxiv} and \textit{no arxiv} conditions. We analyze the aggregate of ICLR 2020 and 2019 papers and also ICLR 2020 papers alone to mitigate for any unintended temporal effects of dataset shift. The p-values from a pair-wise t-test reveal significant differences between the two conditions. Error bars represent $95\%$ confidence intervals.  For each bin, the left plot is the \textit{arxived} condition and the right plot is the \textit{not arxived} condition, as indicated by the legend. }
    \label{fig:unbinned}
\end{figure}

\textbf{Measuring paper reputation}
Since we will be analyzing review outcomes for papers, most of which have multiple authors, we further define a measure for the \textit{pseudo-reputation} of a paper. We will consider the following metrics as definitions of a paper’s pseudo-reputation:
\begin{itemize}
    \item the max of the h-indices/total citations of all authors,
    \item and the average of the h-indices/total citations of the top 2 authors.
\end{itemize}
There are most certainly a number of flaws in using h-index/total citations as a measure of reputation of authors~\cite{hindexbad}, however we intend to clarify that our intention in this work is not to come up with a \textit{better} method of quantifying the reputation of researchers. Given a publicly available standard metric under which research output is quantified (namely h-index and total citation count), our intention is to perform analyses by grouping authors based on this metric, in order to show the existence of variations in acceptance rates of papers under two conditions for different bins of these metrics.

\textbf{Choice of an observational study}  We analyze observational data from ICLR reviews and arXiv. This is \textit{not} a randomized experiment (natural or controlled), so whatever correlations we discover, we will be unable to make strong conclusions about causation. It is possible that a randomized controlled experiment (such as that conducted in ~\cite{singledouble}) would better address an explicitly causal version of our research question. However, we believe the choice of an observational ``counting'' approach strategy is still valuable. Our findings may not generalize well beyond arXiv and ICLR and will be vulnerable to \textit{drift} as publication norms and channels evolve. However, we think there are enough people specifically interested in ICLR and similar machine-learning conferences and that the impact of these conferences is high enough, that this study is still worth pursuing.

\begin{figure*}[t]
    \centering
     \begin{subfigure}[b]{0.5\textwidth}
               \centering
        \includegraphics[width=\columnwidth]{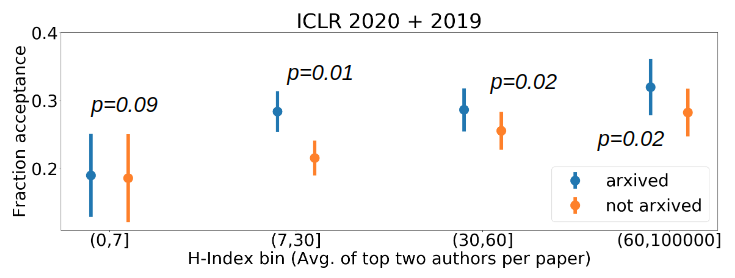}
    \caption{Avg. of top 2 max h indices}
    \label{fig:maintopc}
        \end{subfigure}
           \begin{subfigure}[b]{0.5\textwidth}
               \centering
        \includegraphics[width=\columnwidth]{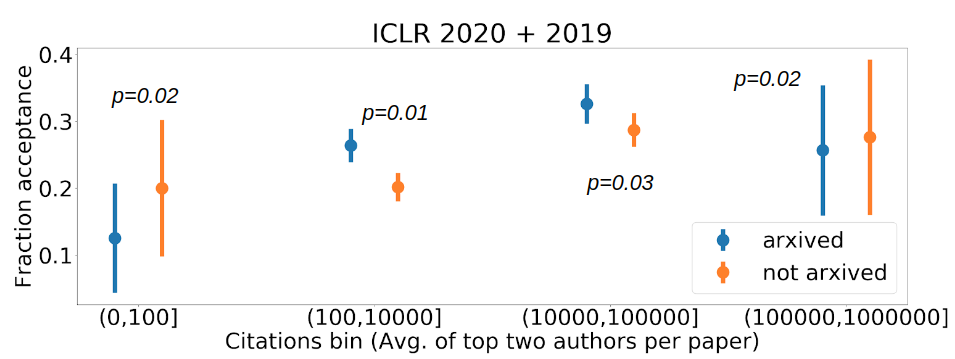}
    \caption{Avg. of top 2 max citations}
    \label{fig:maintop2c}
        \end{subfigure}
      \caption{[ICLR 2020 + ICLR 2019] Analysis of fraction acceptance for different bins of paper pseudo-reputation. The pseudo-reputation metric is the (a) average of the top two max. h indices and the (b) average of the top two max. citations of the author list of each paper. To analyze the statistical significance of our results, we conduct one tailed pairwise t-tests between the two conditions for each of the bins and report the $p$-values in the plots above. Error bars represent $95\%$ confidence intervals.  For each bin, the left plot is the \textit{arxived} condition and the right plot is the \textit{not arxived} condition, as indicated by the legend.  Additional details are present in Section~\ref{sec:main}.}
    \label{fig:main_citation}
\end{figure*}

\begin{figure*}[t]
    \centering
     \begin{subfigure}[b]{0.5\textwidth}
               \centering
        \includegraphics[width=\columnwidth]{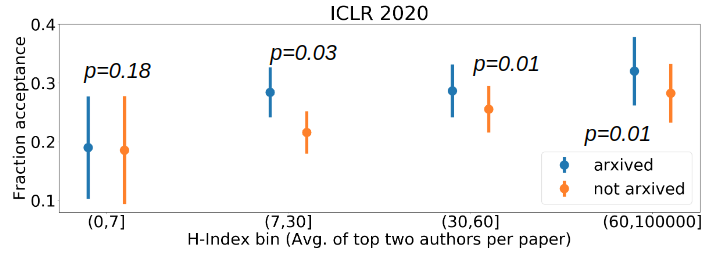}
    \caption{Avg. of top 2 max h indices}
    \label{fig:maintoph}
        \end{subfigure}
           \begin{subfigure}[b]{0.5\textwidth}
               \centering
        \includegraphics[width=\columnwidth]{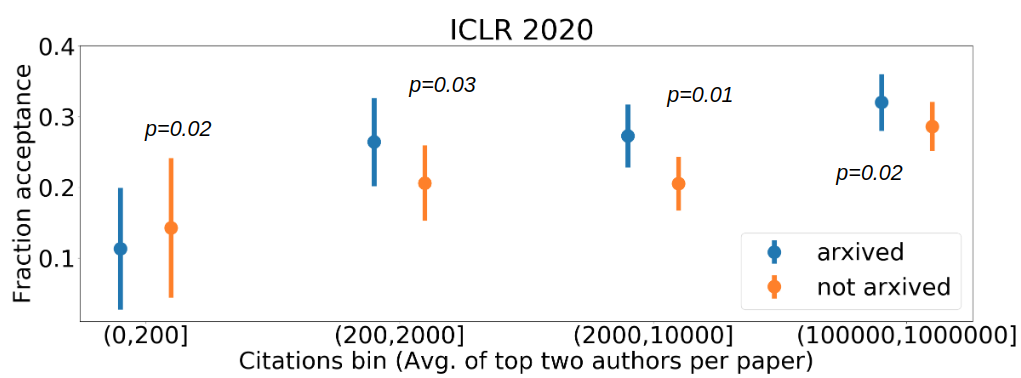}
    \caption{Avg. of top 2 max citations}
    \label{fig:maintop2h}
        \end{subfigure}
      \caption{[ICLR 2020 only] Analysis of fraction acceptance for different bins of paper pseudo-reputation. The pseudo-reputation metric is the (a) average of the top two max. h indices and the (b) average of the top two max. citations of the author list of each paper. To analyze the statistical significance of our results, we conduct one tailed pairwise t-tests between the two conditions for each of the bins and report the $p$-values in the plots above. Here, we analyze ICLR 2020 data only as robustness check for our results to avoid temporal variations in the publication culture from 2019 to 2020 that might confound our analysis. Error bars represent $95\%$ confidence intervals.  For each bin, the left plot is the \textit{arxived} condition and the right plot is the \textit{not arxived} condition, as indicated by the legend.  Additional details are present in Section~\ref{sec:main}.}
    \label{fig:main_hindex}
\end{figure*}


\section{Analyses}
In this section we describe the analyses we performed to understand the research questions. We grouped the analyses under the following headings:

\subsection{Is there any significant difference between acceptance rates for papers that are arxived during/before the review phase and papers that are not?}
We start our analysis by plotting the aggregate acceptance rates of papers in two categories: 1) those whose preprints are released on arxiv either during or before the review phase and 2) the rest whose preprints are either released after the review phase or not released at all till date.

From Fig.~\ref{fig:unbinned} we obtain statistically significant differences between the two conditions. There might be a number of reasons for these differences including the potential explanation that the papers released on arxiv during/before the review typically tend to be more polished than their un-released counterparts. So, we perform additional analyses by binning the \textit{pseudo-reputation} of papers in the subsequent sections to understand the nuances of these differences better. 

\subsection{Do papers with arxiv preprints tend to have higher acceptance rates in case of papers by well-known authors?}
\label{sec:main}
\textbf{\textit{Method:}} We plot a histogram with binned paper pseudo-reputation along the x-axis and average \% of papers accepted in each bin along the y-axis. We consider two different conditions for the plot:
\begin{itemize}
    \item only papers released on arXiv before or during the review process, i.e. before the date reviews were released on OpenReview.
    \item all other papers that are either not present on arXiv or were published on arXiv after the date reviews were released on OpenReview.
\end{itemize}


\textbf{\textit{Results:}} Inspecting the different bins in Fig.~\ref{fig:main_citation}, we note two key trends, 1) the \%acceptance increases with high pseudo-reputation of papers
   in the first bin and 2) the \% acceptance for papers in the \textit{not arxived} condition is higher than the \textit{arxived} condition, while in subsequent bins, in particular the third and fourth bins the trend is reversed.

The first trend aligns with the intuition that papers with high pseudo-reputation have an overall higher \% acceptance rate, because  authors having high author reputation scores perhaps submit genuinely better papers on average. However,this does not explain the second trend of discrepancy we observe between the two conditions. 

To identify if the discrepancies we observe are significant, we conduct pair-wise $t$ tests for the four bins with the null hypothesis $H_0$ being \textit{there is no difference between the \textit{arxiv} and the \textit{no arxiv} conditions.} The alternate hypothesis $H_1$ is that \textit{there is a difference between the \%acceptance in the \textit{arxiv} condition compared to the \%acceptance in the \textit{no arxiv} condition}. Specifically, for the first bin we hypothesize that the \%acceptance in the \textit{arxiv} condition is less than the \%acceptance in the \textit{no arxiv} condition while for the fourth bin we hypothesize that the \%acceptance in the \textit{arxiv} condition is more than the \%acceptance in the \textit{no arxiv} condition.

For the first and fourth bins of Fig.~\ref{fig:maintopc}, we obtain $p=0.09$ and $p=0.02$ respectively. In Fig.~\ref{fig:maintop2c}, we repeat the same analysis, but with h-index of authors used to define the paper pseudo-reputation scores. For the first and fourth bins of Fig.~\ref{fig:maintop2c}, we obtain $p=0.02$ and $p=0.02$ respectively. Hence, we indeed conclude that \textbf{there is a positive correlation between releasing preprints on arXiv and acceptance rates of papers by well-known authors}, under our concretization of the problem.

Since the data for Fig.~\ref{fig:main_citation} consists of ICLR 2020 and ICLR 2019 papers combined, in order to ensure that our results are not confounded by temporal changes in the publication culture over one year, we perform a robustness check by repeating the analyses for ICLR 2020 papers alone in Fig.~\ref{fig:main_hindex}. For the first and fourth bins of Fig.~\ref{fig:maintoph}, we obtain $p=0.18$ and $p=0.01$ respectively, while for the first and fourth bins of Fig.~\ref{fig:maintop2c}, we obtain $p=0.02$ and $p=0.02$ respectively. These results are consistent with those in Fig.~\ref{fig:main_citation} and hence our conclusions remain valid.

\subsection{Are high pseudo-reputation papers more likely to be released on arXiv?}
\label{sec:frac}

We hypothesize that papers with high pseudo-reputation are more likely to have preprints released on arXiv during or before review. 
Authors understandably want the best outcome for their paper especially when they understand the amount of time, effort, and analysis that has gone into their papers. So, it may be the case that well known authors believe deblinding via arxiv and publicizing their paper before/during peer review will likely work in their favor.

\textbf{\textit{Method:}} To test this hypothesis we plot a histogram with binned paper pseudo-reputation along the x-axis and fraction of papers released on arXiv as y-axis. Note that there is only a single condition per bin in this histogram unlike the previous plots, and we are interested in comparing the y-values corresponding to each bin.

\begin{figure}
    \centering\hspace*{-0.5cm}
    \includegraphics[width=1.1\columnwidth]{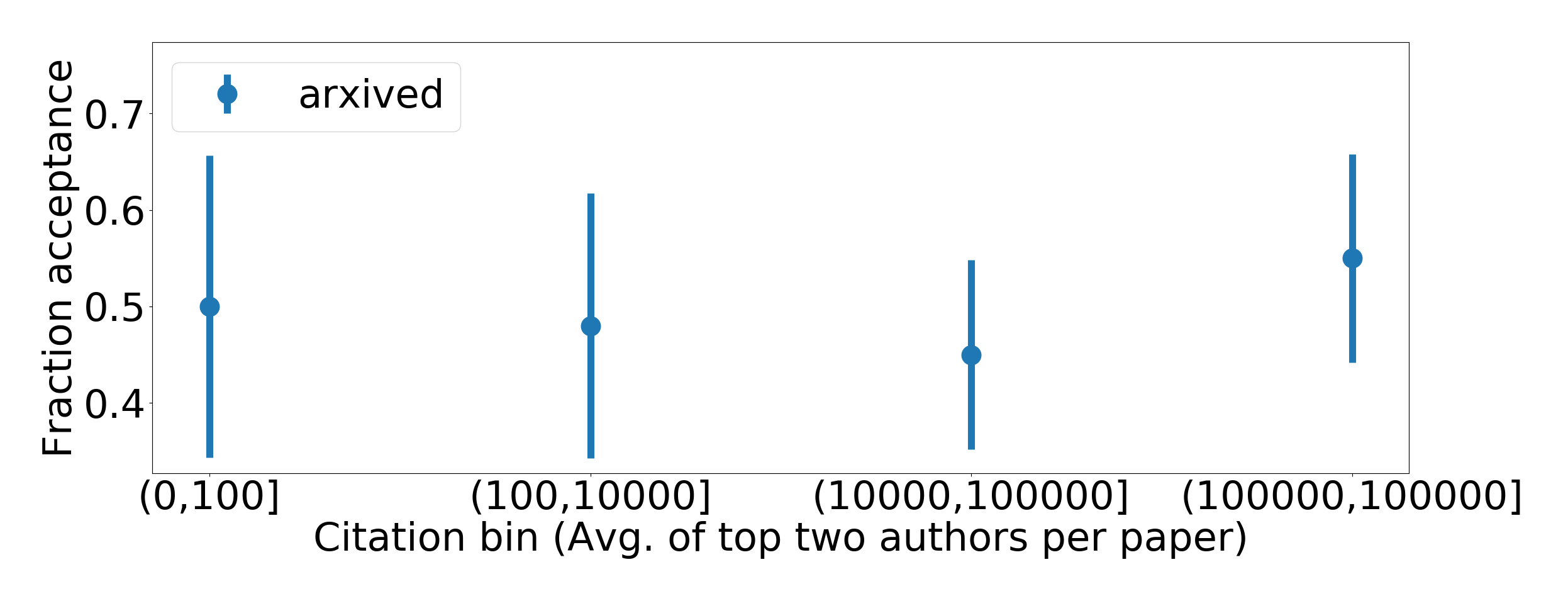}
    \caption{\%of papers released on arXiv during or before the review phase, with respect to the number of submissions in each citation bin. The citation bins are computed with the paper pseudo-reputation metric of the avg. of top 2 cited authors in each paper. Error bars represent $95\%$ confidence intervals. Detailed analysis (including p-tests that take S.D. into account)  is present in Section~\ref{sec:frac}.}
    \label{fig:frac_arxived}
\end{figure}

\textbf{\textit{Results:}} The result of this analysis is shown in Fig.~\ref{fig:frac_arxived}. While it is evident that the fraction of papers arxived in the fourth category is more than all the other three categories, the difference at least through visual inspection is not profound. To quantify if there is a statistically significant difference of $\%$papers arxived in the case of \textit{high} pseudo-reputation of papers and \textit{low} pseudo-reputation of papers, we perform a $t$ test on the aggregate of the first three bins (the \textit{low} condition) and the fourth bin (the \textit{high} condition), with the alternate hypothesis $H_2$ that \textit{the $\%$papers arxived is higher in the \textit{high} category compared to the \textit{low} category}. The null hypothesis $H'_0$ is that \textit{there is no difference in $\%$papers arxived for the two conditions \textit{high} and \textit{low}}. For this, we obtain $p=0.04$, which does allow us to reject the null hypothesis in favor of the alternate hypothesis. 

\begin{figure*}[t]
    \centering
     \begin{subfigure}[b]{0.5\textwidth}
              \centering
        \includegraphics[width=\columnwidth]{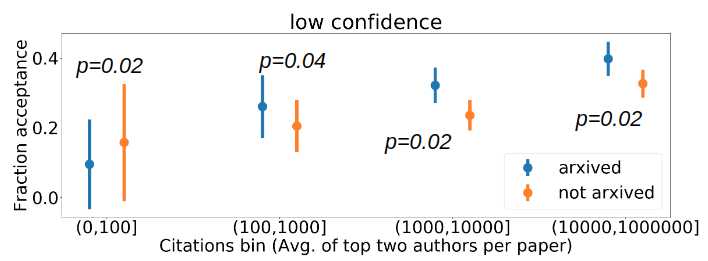}
    \caption{low confidence (citations)}
    \label{fig:low}
        \end{subfigure}\hspace*{-0.1em}%
          \begin{subfigure}[b]{0.5\textwidth}
              \centering
        \includegraphics[width=\columnwidth]{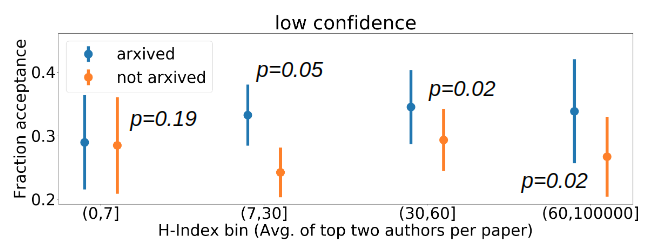}
    \caption{low confidence (h-index)}
    \label{fig:hindlow}
        \end{subfigure}\hspace*{-0.1em}%

          \begin{subfigure}[b]{0.5\textwidth}
              \centering
        \includegraphics[width=\columnwidth]{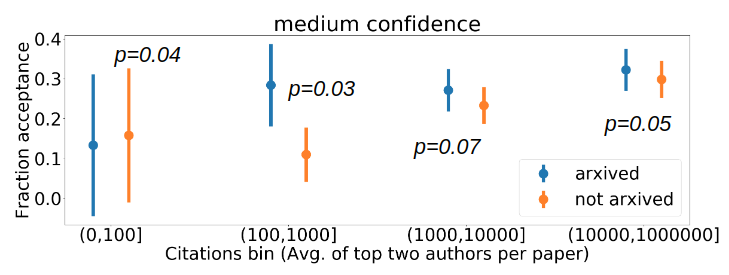}
    \caption{medium confidence (citations)}
    \label{fig:medium}
        \end{subfigure}\hspace*{-0.1em}%
         \begin{subfigure}[b]{0.5\textwidth}
              \centering
        \includegraphics[width=\columnwidth]{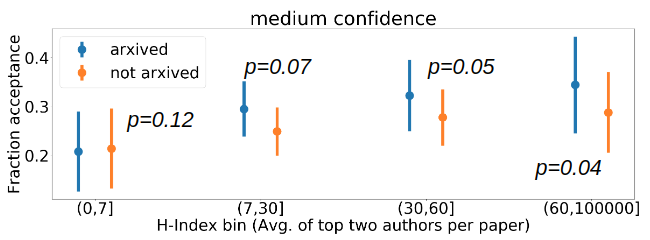}
    \caption{medium confidence (h-index)}
    \label{fig:hindmedium}
        \end{subfigure}\hspace*{-0.1em}%

         \begin{subfigure}[b]{0.5\textwidth}
              \centering
        \includegraphics[width=\columnwidth]{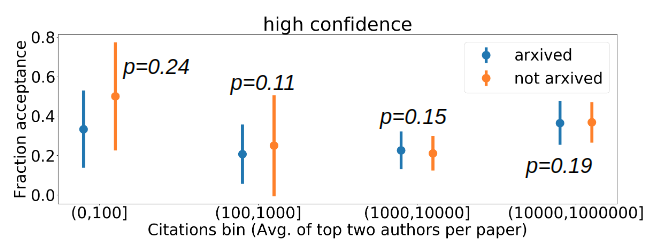}
    \caption{high confidence (citations)}
    \label{fig:high}
        \end{subfigure}
         \begin{subfigure}[b]{0.5\textwidth}
              \centering
        \includegraphics[width=\columnwidth]{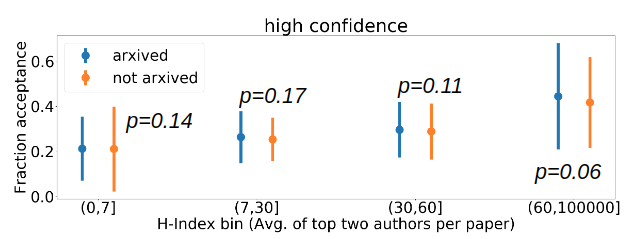}
    \caption{high confidence (h-index)}
    \label{fig:hindhigh}
        \end{subfigure}
      \caption{Analysis of the percent acceptance with respect to paper pseudo-reputation citation bins for different reviewer confidence scores. For the left column, the pseudo-reputation metric is the max. citation of the author list of each paper and the average reviewer confidence scores are grouped into (a) low, (c) medium, and (e) high categories. For the right column, the pseudo-reputation metric is the average of the top two h-indices of the author list of each paper and the average reviewer confidence scores are grouped into (b) low, (d) medium, and (f) high categories. To analyze the statistical significance of our results, we conduct one tailed pairwise t-tests between the two conditions for each of the bins and report the $p$-values in the plots above. Error bars represent $95\%$ confidence intervals. For each bin, the left plot is the \textit{arxived} condition and the right plot is the \textit{not arxived} condition, as indicated by the legend. Additional details are present in Section~\ref{sec:confidence}.}
    \label{fig:confidence}
\end{figure*}

\subsection{Are review scores by \textit{less confident} reviewers higher in case of papers with high pseudo-reputation and lower in case of papers with low pseudo-reputation?}
\label{sec:confidence}

While writing reviews for ICLR papers, reviewers must self-specify their \textit{confidence} in the review of the paper in the form of a field called \textit{experience assessment}. There are four different confidence levels that reviewers can choose from, for example the highest level is defined by ``I have read many papers in this area." This is publicly displayed along with the reviews. We denote the numerical value of the confidence scores as $1,2,3,4$ (lowest to highest in this order).
\begin{figure}[t]
               \centering
       \hspace*{-0.5cm} \includegraphics[width=1.1\columnwidth]{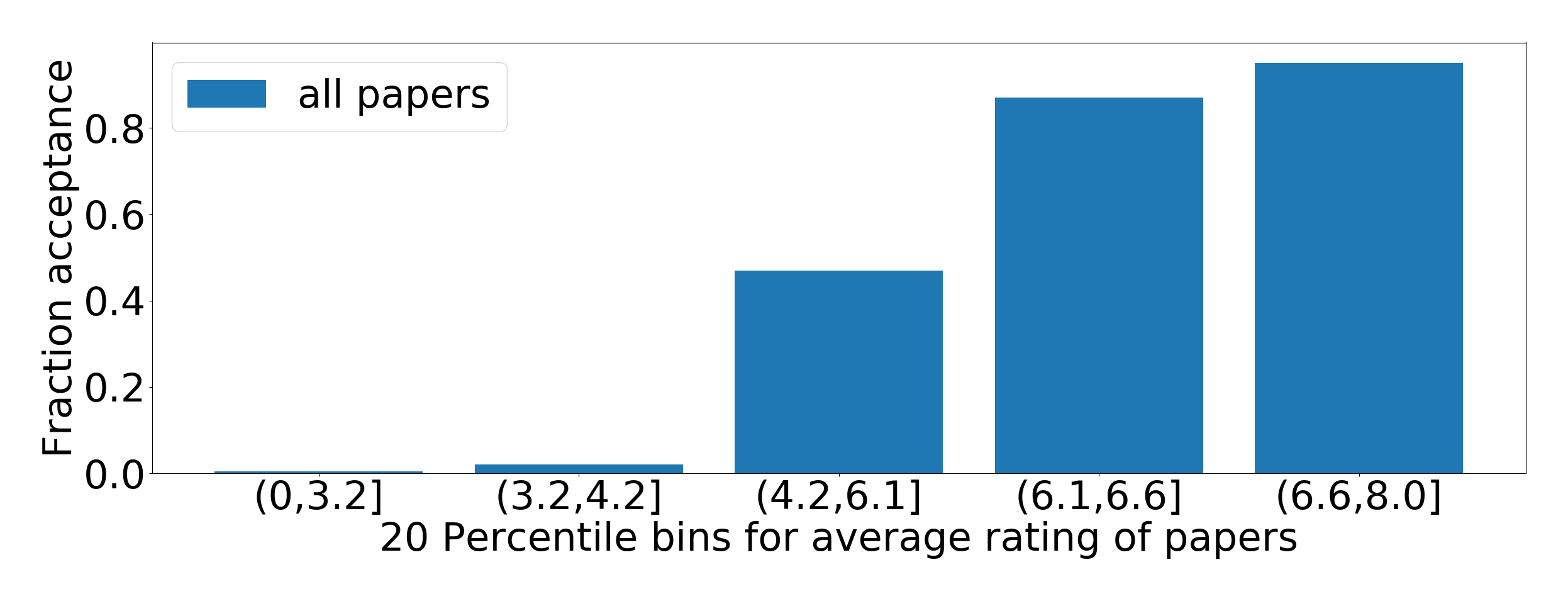}
    \caption{Plot of fraction acceptances of papers in each of the twenty percentile bins of average paper ratings. We consider this plot to identify as suitable range for defining \textit{borderline} and \textit{highly rated} papers. Since fraction acceptance in the $(4.2,6.1]$ bin is roughly $0.5$, we define all papers with average reviewer ratings in this range to be \textit{borderline} for the analysis in Section~\ref{sec:borderline}. }
    \label{fig:borderlinepercentile}
        \end{figure}

\begin{figure*}
    \centering 
     \begin{subfigure}[b]{0.5\textwidth}
               \centering
        \includegraphics[width=\columnwidth]{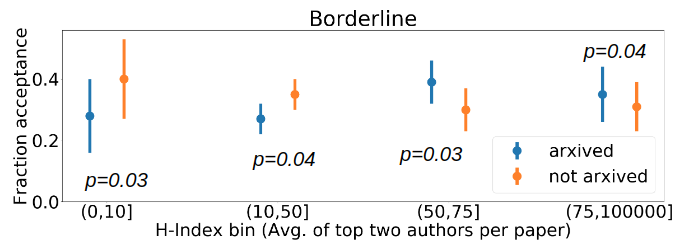}
    \caption{Borderline [4.2,6.1]}
    \label{fig:borderline}
        \end{subfigure}
           \begin{subfigure}[b]{0.5\textwidth}
               \centering
        \includegraphics[width=\columnwidth]{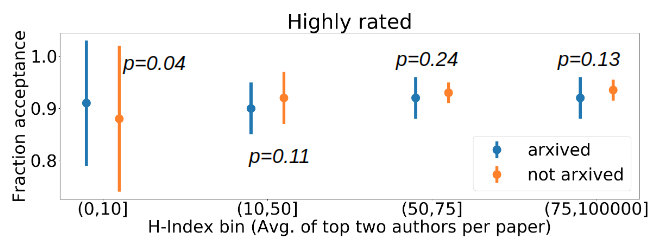}
    \caption{Highly rated [6.1,8]}
    \label{fig:highrated}
        \end{subfigure}
      \caption{Analysis of \%acceptance for different H-Index bins of paper pseudo-reputation. The pseudo-reputation metric is the average of the top two max. h-indices of the author list of each paper. In (a) we analyze borderline papers and in (b) we analyze highly rated papers. To analyze the statistical significance of our results, we conduct one tailed pairwise t-tests between the two conditions for each of the bins and report the $p$-values in the plots above. Error bars represent $95\%$ confidence intervals. For each bin, the left plot is the \textit{arxived} condition and the right plot is the \textit{not arxived} condition, as indicated by the legend. Additional details are present in Section~\ref{sec:borderline}.}
    \label{fig:borderline_analysis}
\end{figure*}

\textbf{\textit{Method:}} We consider bins of paper pseudo-reputations on the x-axis for all papers that have been released on arXiv and plot 3 histograms (corresponding to whether the average reviewer confidence score for the paper lies in $low$ [1,2.5], $medium$ (2.5,3], or $high$ (3,4] categories) indicating the \textit{average} review score assigned by each category of reviewers to papers in each bin. 

\textbf{\textit{Results:}} Fig.~\ref{fig:confidence} shows the results of this analysis. Looking at the third and fourth bins in Fig.~\ref{fig:low}, Fig.~\ref{fig:medium}, and Fig.~\ref{fig:high}, it is evident that for papers with a low average reviewer confidence score, the average review score in the \textit{arxiv} condition is more than the average review score in the \textit{no arxiv} condition. Looking at the fist bin in Fig.~\ref{fig:low}, Fig.~\ref{fig:medium}, and Fig.~\ref{fig:high}, it is evident that for papers with a low average reviewer confidence score, the average review score in the \textit{arxiv} condition is less than the average review score in the \textit{no arxiv} condition. 

To analyze if these differences are significant, we conduct $t$ tests on the four bins. The null hypothesis $H''_0$ is that \textit{there is no difference between the \textit{arxiv} and the \textit{no arxiv} conditions.} The alternate hypothesis $H_3$ is that for low confidence reviewers, \textit{there is a difference between the \textit{arxiv} and the \textit{no arxiv} conditions.} Specifically, we hypothesize that in the third and fourth bins, the average review score in the \textit{arxiv} condition is more than the average review score in the \textit{no arxiv} condition. We also hypothesize that in the first bin, the average review score in the \textit{arxiv} condition is less than the average review score in the \textit{no arxiv} condition.

For the fourth bins in Fig.~\ref{fig:low}, Fig.~\ref{fig:medium}, and Fig.~\ref{fig:high}, we respectively obtain the $p$ values 0.02, 0.03, and 0.19. Since $p\leq0.05$ for Fig.~\ref{fig:low} and Fig.~\ref{fig:medium}, we can reject the null hypothesis in favor of the alternate hypothesis, but we cannot reject the null hypothesis for Fig.~\ref{fig:high}. This offers \textbf{evidence of negative correlation between \textit{confidence} of reviewers and their likelihood to assign high review scores to papers with high pseudo-reputation}.

For the first bin in Fig.~\ref{fig:low}, Fig.~\ref{fig:medium}, and Fig.~\ref{fig:high}, we respectively obtain the $p$ values 0.02, 0.04, and 0.24. Since $p\leq0.05$ for Fig.~\ref{fig:low}, we can reject the null hypothesis in favor of the alternate hypothesis, but we cannot reject the null hypothesis for Fig.~\ref{fig:medium}, and Fig.~\ref{fig:high}. This offers \textbf{evidence of negative correlation between \textit{confidence} of reviewers and their likelihood to assign low review scores to papers with low pseudo-reputation}.

We repeat our analyses with average of top two authors' h-indices as the metric for paper pseudo-reputation in Figs.~\ref{fig:hindhigh},~\ref{fig:hindmedium}, and~\ref{fig:hindlow} and obtain similar conclusions

\subsection{Is the effect of difference between the \textit{arxiv} and the \textit{no arxiv} conditions stronger for \textit{borderline} papers?}
\label{sec:borderline}
In Fig.~\ref{fig:borderline_analysis} we analyze papers that have a borderline reviewer rating on average and papers that are highly rated by reviewers on average, under the two conditions \textit{arxived} and \textit{not arxived} prior to decision notification. This analysis aims to understand the existence of potential bias at the level of Area Chairs.

\noindent \textbf{\textit{Method:}} To have a principled scheme of determining which papers are \textit{borderline}, in Fig.~\ref{fig:borderlinepercentile}, we plot fraction acceptance of papers per average reviewer rating bin, where the bins are created based on the twenty percentile values. Based on this, we define \textit{borderline} papers to be the papers that received an average rating in the range $[4.2,6.1]$ and \textit{highly rated} papers to be those that received an average rating in the range $[6.1,8]$. 

\noindent \textbf{\textit{Results:}} Fig.~\ref{fig:borderline} and Fig.~\ref{fig:highrated} respectively show fraction acceptance per h-index bin for the borderline papers and the highly rated papers. To identify if the discrepancies we observe are significant, we conduct $t$ tests for the four bins with the null hypothesis $H'''_0$ being \textit{there is no difference between the \textit{arxiv} and the \textit{no arxiv} conditions.} The alternate hypothesis $H_4$ is that \textit{there is a difference between the \textit{arxiv} and the \textit{no arxiv} conditions.} Specifically, for the first bin, we hypothesize that the \%acceptance in the \textit{arxiv} condition is less than the \%acceptance in the \textit{no arxiv} condition, while for the fourth bin, we hypothesize that the \%acceptance in the \textit{arxiv} condition is more than the \%acceptance in the \textit{no arxiv} condition.

For the first and fourth bins of Fig.~\ref{fig:borderline}, we obtain $p=0.03$ and $p=0.03$ respectively, while for the first and fourth bins of Fig.~\ref{fig:highrated}, we obtain $p=0.04$ and $p=0.13$ respectively. So for the borderline papers, we conclude that releasing preprints on arXiv correlates positively with acceptance rates of papers by well-known authors, and correlates negatively with acceptance rates of papers by less well known authors under our concretization of the problem. \textbf{ The effect is indeed stronger for borderline papers}.

\section{Discussion and Limitations}
In the previous section we performed a number of analyses and obtained three key inferences 1) releasing preprints on arXiv has a positive correlation with acceptance rates of papers by well-known authors, 2) papers with well-known authors are more likely to be released on arXiv during or prior to the review phase, and 3) reviewers with a low confidence score are more likely to assign high review scores to de-anonymized papers by well-known authors. In this section we intend to put these inferences in the right perspective and address some of the limitations of our study. 

It is important to note that our study is entirely based on observational data and hence it is not possible for us to make rigorous causal claims. Since the same set of reviewers were not exposed to the two conditions \textit{arxiv} and \textit{no arxiv} we cannot make any conclusive claims with respect to the \textit{intent} or \textit{bias} of the reviewers. On the other hand, we believe that the in-the-wild nature of our study is helpful in putting into perspective the trends that emerge (albeit correlational and not necessarily causal)  in the current publication and preprint culture of machine learning. 

Another limitation of our study is that we only analyze data from two recent ICLR conferences (ICLR 2020 and ICLR 2019). ICLR served as the natural platform for this study as the entire list of submissions and reviews are publicly released, in the spirit of open science. It would be very helpful if we could validate our claims on other popular CS/AI/ML conferences to understand the interplay of de-anonymization through arXiv and the type of reviews. This is our appeal to the community to consider adopting the OpenReview system and publicly release the entire list of submissions and all the reviews. Apart from facilitating analyses like ours, this also helps readers put into perspective the contributions of the papers and understand the potential shortcomings that were pointed out in the review phase and that \textit{hopefully} have been addressed in the final version. 

Finally, given our findings in this study and the implications this must be having in the publication culture of our community, we discuss some solutions to mitigate potential reviewer bias caused by de-anonymization through arXiv preprints. Since the point of a preprint is that the paper is either soon to be submitted for review or is currently under review, arxiv.org could have the option of allowing authors to keep the author list anonymized. Conferences that follow the double blind review system could enforce that only papers that are anonymized on arXiv and that will remain anonymized during the review phase can be submitted to the conference. If the purpose of releasing a preprint is early dissemination of knowledge, having the author list anonymized for some duration would not be detrimental to this cause as the anonymized paper would still be citable (a practice followed by ICLR on OpenReview).  In order to strongly discourage people from putting up incomplete works under the cover of anonymity for `early flagplanting,' strict rules can be enforced regarding the conditions under which a paper submitted on arxiv can be later updated. For example, it can be imposed that papers that are submitted in anonymous format when updated will still display the old version \textbf{as default} and the new version will have a \textbf{separate} upload timestamp and be linked to the old version. i.e. the two versions (anonymous and updated) will be listed as separate papers with their respective timestamps in order to discourage people from trying to flagplant incomplete work.

In addition to the above, we believe it is important to modify the typical peer-review process to have a maximum limit on the number of \textit{low confidence} reviewers that are assigned to a paper. Since we have observed correlational evidence in that low confidence reviewers are more likely to assign favorable ratings to papers with reputable authors, if possible the number of \textit{low confidence} reviewers overall in the review process should be decreased and if this is not possible then the number of such reviewers per paper must be limited to atmost one.


\section*{Acknowledgements}
We thank the numerous people who pursue meta-discussions about the current publication culture on social media platforms, in particular on Twitter. Those discussions served as an inspiration for us to pursue a principled study to investigate the correlations between reputation of authors and de-blinding of their papers through arxiv submissions prior to double-blind peer review. We thank David Duvenaud and Florian Shkurti for helpful discussions and feedback.

\bibliographystyle{ACM-Reference-Format}
\bibliography{sample-base}


\end{document}